\documentclass[prl,twocolumn,superscriptaddress,showpacs,epsf]{revtex4}

\usepackage{graphicx}
\usepackage{latexsym}
\usepackage{amsmath}
\usepackage{amssymb}
\usepackage{amsfonts}
\usepackage{color}

\begin{document}

\title{Symmetry-induced giant vortex state in a superconducting film with a Penrose array of magnetic pinning centers}


\author{R. B. G. Kramer}
\affiliation{INPAC-Institute for Nanoscale Physics and Chemistry, Nanoscale Superconductivity and Magnetism $\&$ Pulsed Fields Group, K. U. Leuven Celestijnenlaan 200 D, B-3001 Leuven, Belgium.}

\author{A. V. Silhanek}
\affiliation{INPAC-Institute for Nanoscale Physics and Chemistry, Nanoscale Superconductivity and Magnetism $\&$ Pulsed Fields Group, K. U. Leuven Celestijnenlaan 200 D, B-3001 Leuven, Belgium.}

\author{J. Van de Vondel}
\affiliation{INPAC-Institute for Nanoscale Physics and Chemistry, Nanoscale Superconductivity and Magnetism $\&$ Pulsed Fields Group, K. U. Leuven Celestijnenlaan 200 D, B-3001 Leuven, Belgium.}

\author{B. Raes}
\affiliation{INPAC-Institute for Nanoscale Physics and Chemistry, Nanoscale Superconductivity and Magnetism $\&$ Pulsed Fields Group, K. U. Leuven Celestijnenlaan 200 D, B-3001 Leuven, Belgium.}


\author{V. V. Moshchalkov}
\affiliation{INPAC-Institute for Nanoscale Physics and Chemistry, Nanoscale Superconductivity and Magnetism $\&$ Pulsed Fields Group, K. U. Leuven Celestijnenlaan 200 D, B-3001 Leuven, Belgium.}


\date{\today}
\begin{abstract}

A direct visualization of the flux distribution in a Pb film covering a five-fold Penrose array of Co dots is obtained by mapping the local field distribution with a scanning Hall probe microscope. We demonstrate that stable vortex configurations can be found for fields $H \sim 0.8 H_1$, $H_1$ and $1.6 H_1$, where $H_1$ corresponds to one flux quantum per pinning site. The vortex
pattern at $0.8 H_1$ corresponds to one vacancy in one of the vertices of the thin tiles whereas at $1.6 H_1$ the vortex structure can be associated with one interstitial vortex inside each thick tile. Strikingly, for $H=1.6 H_1$ interstitial and pinned vortices arrange themselves in ring-like structures (``vortex corrals") which favor the formation of a giant vortex state at their center. 

\end{abstract}


\maketitle

Crystals are classified according to their translational and rotational symmetry. Until mid 80's, the lack of crystalline structures having a five fold rotation symmetry axis was generally accepted. However, the revolutionary discovery in 1984 of crystals with ``forbidden" symmetry~\cite{shechtman-PRL-94}, introduced a new family of crystallographic structures, known as quasicrystals, with order but not periodic. The existence of quasiperiodic arrays in two dimensions was earlier introduced by Penrose~\cite{penrose} in a mathematical context, who demonstrated that a two-dimensional plane can be fully covered by combining two different unit cells or tiles. The resulting structures possess long range orientational order without having long range translational order.

The interest for the unique properties of quasicrystals has nowadays surpassed the boundaries of the crystallographic community and reached most research fields in condensed matter physics~\cite{nori-PRB-1986,kohmoto-PRB-1987,springer-PRL-1987,nelson-halperin,abe-nature-2004}. Particular attention has been recently devoted to the pinning properties of flux lines in superconducting thin films with aperiodic, fractal or quasiperiodic arrays of pinning sites~\cite{misko-PRL,misko-PRB,villegas,kemmler,silhanek,reichhardt}. In this case, the lack of translational order of the pinning landscape prevents the formation of one-dimensional channels for easy vortex flow whereas the convolution of many build-in periods favors the proliferation of matching effects. In principle, both properties tend to improve the maximum current attainable without dissipation, i.e. the superconducting critical current.

Molecular dynamic simulations of driven vortices interacting with a five-fold two dimensional Penrose array of pinning sites predicted local enhancements of the critical current for external fields $H = 0.757 H_1$, $H_1$, and $1.482 H_1$, where $H_1$ is the field at which the density of singly quantized vortices coincides with the density of pinning centers~\cite{misko-PRL,misko-PRB}. The non trivial matching features below and above $H_1$, should correspond to the occupancy of three out of four pinning sites in the vertices of the thin tiles and the presence of an interstitial vortex in each thick tile, respectively. Indirect evidence of these stable vortex configurations has been obtained by transport measurements for a Penrose array of holes in a Nb film~\cite{kemmler} and magnetic dots in Al and Pb films~\cite{silhanek}. In both experimental investigations further unforseen features were reported. Unfortunately, the lack of direct visualization of the vortex distribution in this sort of systems have concealed the real space vortex arrangement associated with the observed matching features in the transport properties.


In this work we establish a clear correlation between the theoretical predictions and the transport measurements by mapping the local field distribution in a Pb film covering a Penrose array of Co dots via scanning Hall probe microscopy (SHPM). The vortex patterns directly visualized at $H = 0.765 H_1$ and $H = H_1$ are in agreements with the theoretical expectations. For $H > H_1$, interstitial vortices are placed inside the thick tiles, as anticipated by molecular dynamic simulations~\cite{misko-PRL,misko-PRB}. However, these interstitial vortices lie in a bistable state rather than in the geometrical center of the tile. Due to the long range vortex-vortex interaction, this degeneracy is lifted by the lack of reflection symmetry of the surrounding tiles. As a result, at certain magnetic fields the interstitial vortices accommodate themselves in order to achieve an ordered ring-like structure involving several tiles. Strikingly, this highly symmetric ring like structure can stabilize a giant vortex state of two flux quanta at the center of the ring. This new symmetry-induced giant vortex state reminds that obtained in disk-shaped mesoscopic structures and results from the local field pressure exerted by the surrounding vortices. Complementary transport measurements show that this behavior leads to a clear matching feature at $H = 1.6 H_1$. Further unforeseen commensurability effects are also found at $H = 3.8 H_1$ and $H = 6.0 H_1$.

\begin{figure}[bt]
\centering
\includegraphics*[width=0.7\linewidth, bb=43 41 293 293]{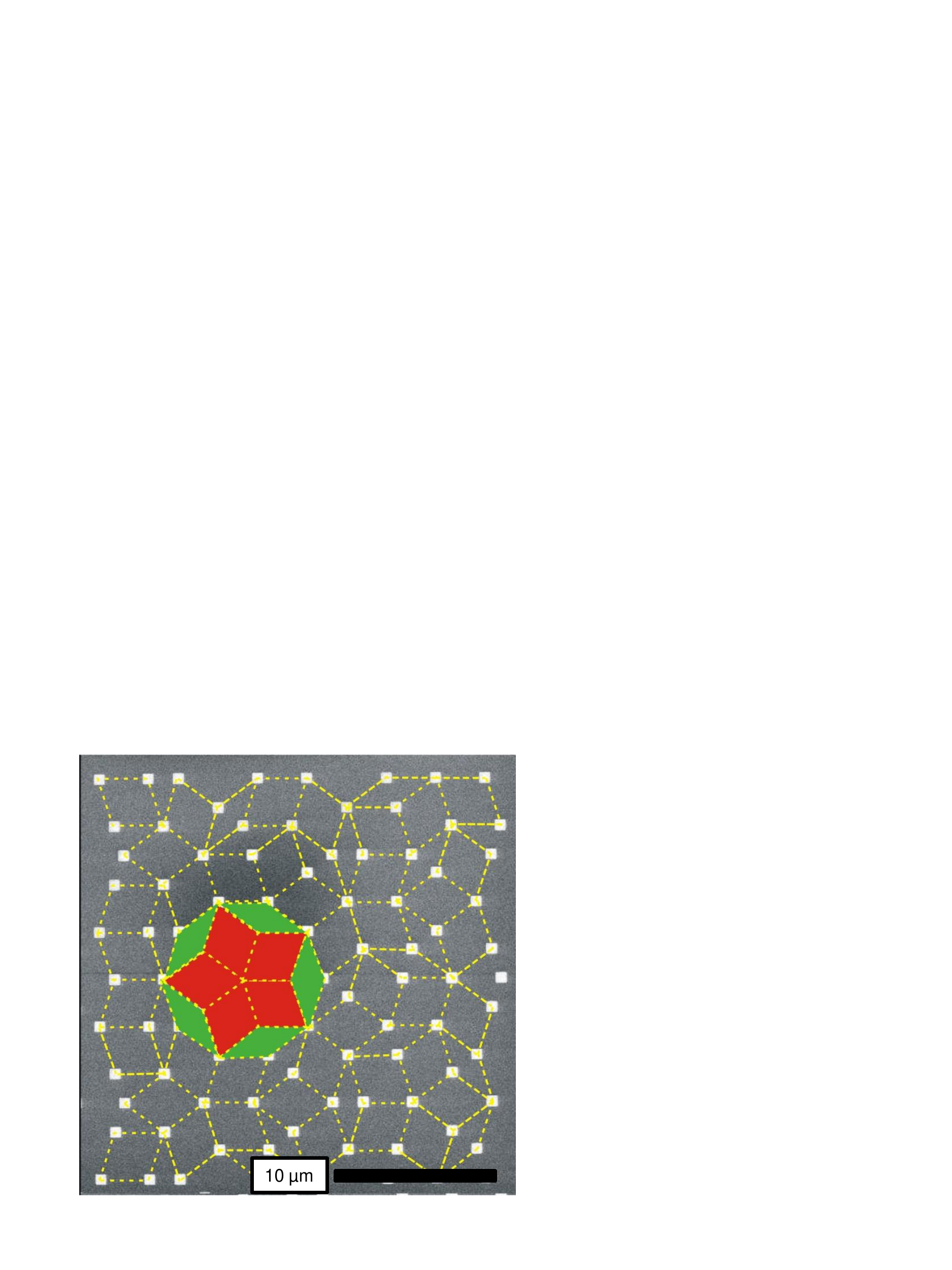}
\caption{(color online) Scanning electron microscopy image of the square dots distributed in a five-fold Penrose array. For clarity, the two types of tiles are indicated with dotted (yellow) lines. Five thin and thick tiles are painted with green and red color, respectively. The black bar indicates a distance of 10~$\mu$m.}
  \label{SEM}
\end{figure}

The investigated samples consist of a 50~nm thick Pb film evaporated directly on top of a five-fold Penrose array of square Co dots made by electron beam lithography and subsequent lift-off. The experimental procedure used for the sample preparation can be found in Ref.~\onlinecite{samplepreparation}. Figure~\ref{SEM} shows a scanning electron microscopy image of the dots' array. For clarity the magnetic particles have been connected by dotted lines to indicate the distribution of the thin and thick tiles, which represent the building blocks of the five-fold Penrose lattice. The size of the Co dots is 0.7~$\mu$m and the length of the connecting lines is 3.1~$\mu$m. The pinning potential is created predominantly by a local depletion of superconductivity above the Co dots due to the proximity effect at the Co/Pb interface. Experiments performed before and after magnetizing the dots, show no differences, thus confirming that the electromagnetic coupling does not play a relevant role. For transport measurements the samples were patterned in a bridge shape 102~$\mu$m wide with a voltage contacts' separation of 392~$\mu$m. The Pb film has a critical temperature $T_{c0}$=7.238~K as determined by the 50$\%$ of the normal state resistance.

The scanning Hall probe microscopy images were obtained using a modified low temperature scanning Hall probe microscope (SHPM) from Nanomagnetics Instruments~\cite{Bending}. Most of the SHPM images shown in this work were recorded at 4.2~K with a scanning area of $14 \times 14~\mu$m$^2$ after field cooling  the sample. Measurements performed at higher temperatures show no difference with those obtained at 4.2~K, thus indicating that the vortex distribution is frozen at high temperatures close to the onset of the superconducting state. The images were recorded in lift-off mode with the Hall sensor at about 500~nm above the surface of the sample. Home-made $xy$ positioners allow us to explore different regions of the same sample in order to avoid unwanted effects arising from the sample borders, imperfections, or small particles.

\begin{figure}[bt]
\centering
\includegraphics*[width=\linewidth, bb=0 0 430 410]{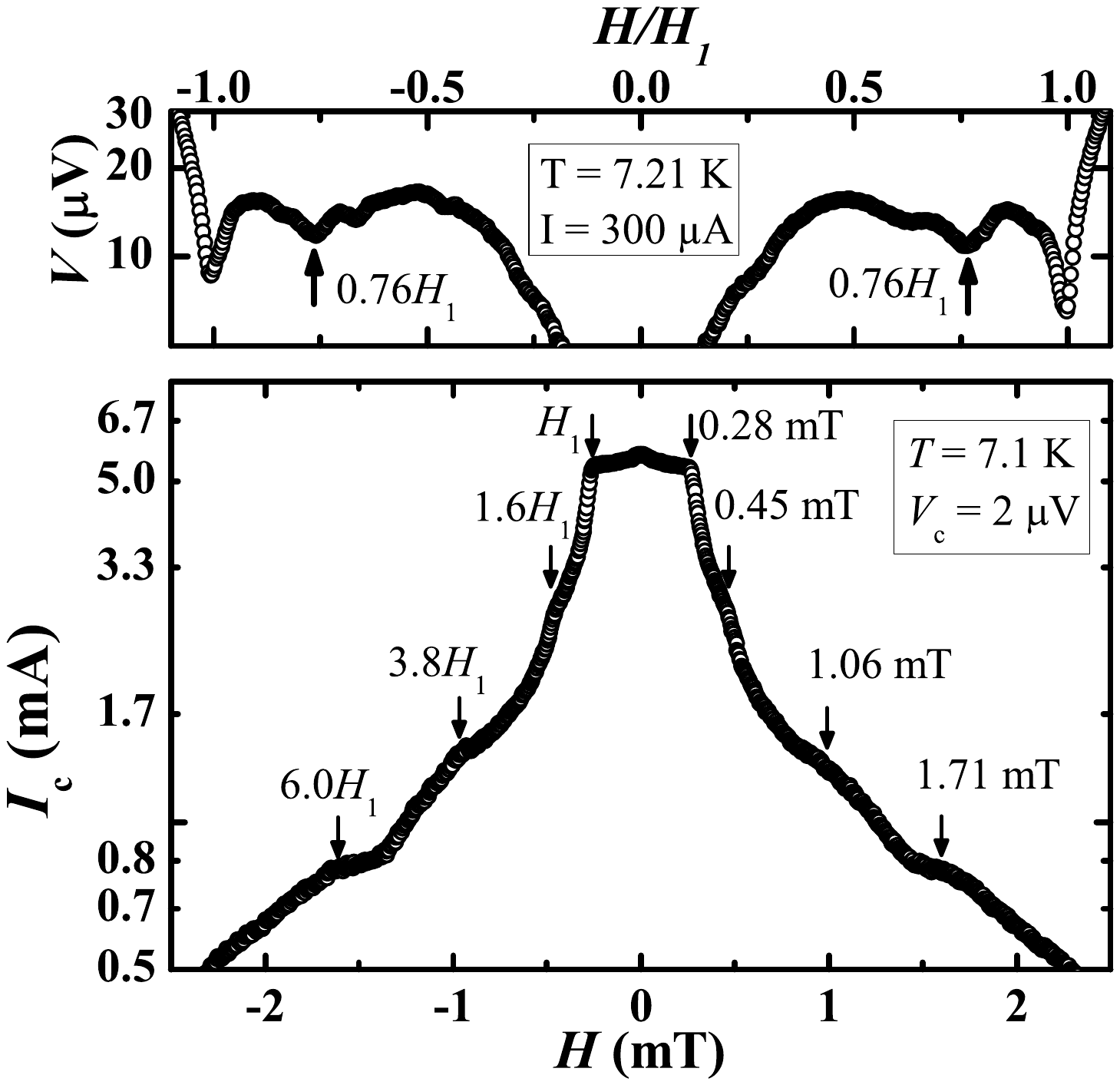}
\caption{The upper panel shows the voltage for a fixed bias current of 300~$\mu$V as a function of the external field at $T=7.21$~K. The lower panel shows the superconducting critical current, estimated with a voltage criterion of 2~$\mu$V, as a function of the external field for $T=7.1$~K. Matching features are indicated with black arrows. } \label{transport}
\end{figure}

We first investigated the vortex dynamics of the superconducting sample by performing a series of current-voltage characteristics as a function of field $H$ at temperatures very close to $T_{c0}$. A detailed measurement of the voltage $V$ vs $H$ at constant current $I=300~\mu$A and $T=7.21$~K is shown in the upper panel of Fig.~\ref{transport}. The first matching field $H_1 \sim 0.28$~mT can be clearly identified as a pronounced dip in the dissipation, indicating that each pinning site can trap a maximum of one flux quantum. An additional dip is present at $H \sim 0.76 H_1$ as predicted by molecular dynamic simulations. As we demonstrate below this field corresponds to a vortex distribution leaving one vacancy in one of the vertices of the thin tiles.

To show the presence of stable vortex configurations at higher densities we plot the depinning current $I_c$, using a dissipation criterion of 2~$\mu$V, for $T=7.1$~K (Fig.~\ref{transport}, lower panel). Higher matching features can be seen for $H \sim 1.6 H_1$, $3.8 H_1$, and $6.0 H_1$. These relatively high order matching features point out the relevance of the orientational order in stabilizing the vortex lattice. It it worth mentioning that typically commensurability effects in periodic pinning arrays are attributed to the perfect compensation of vortex currents, a condition which seems to be never satisfied in a quasiperiodic array.

\begin{figure}[bt]
\centering
\includegraphics*[width=\linewidth, bb=29 129 480 540]{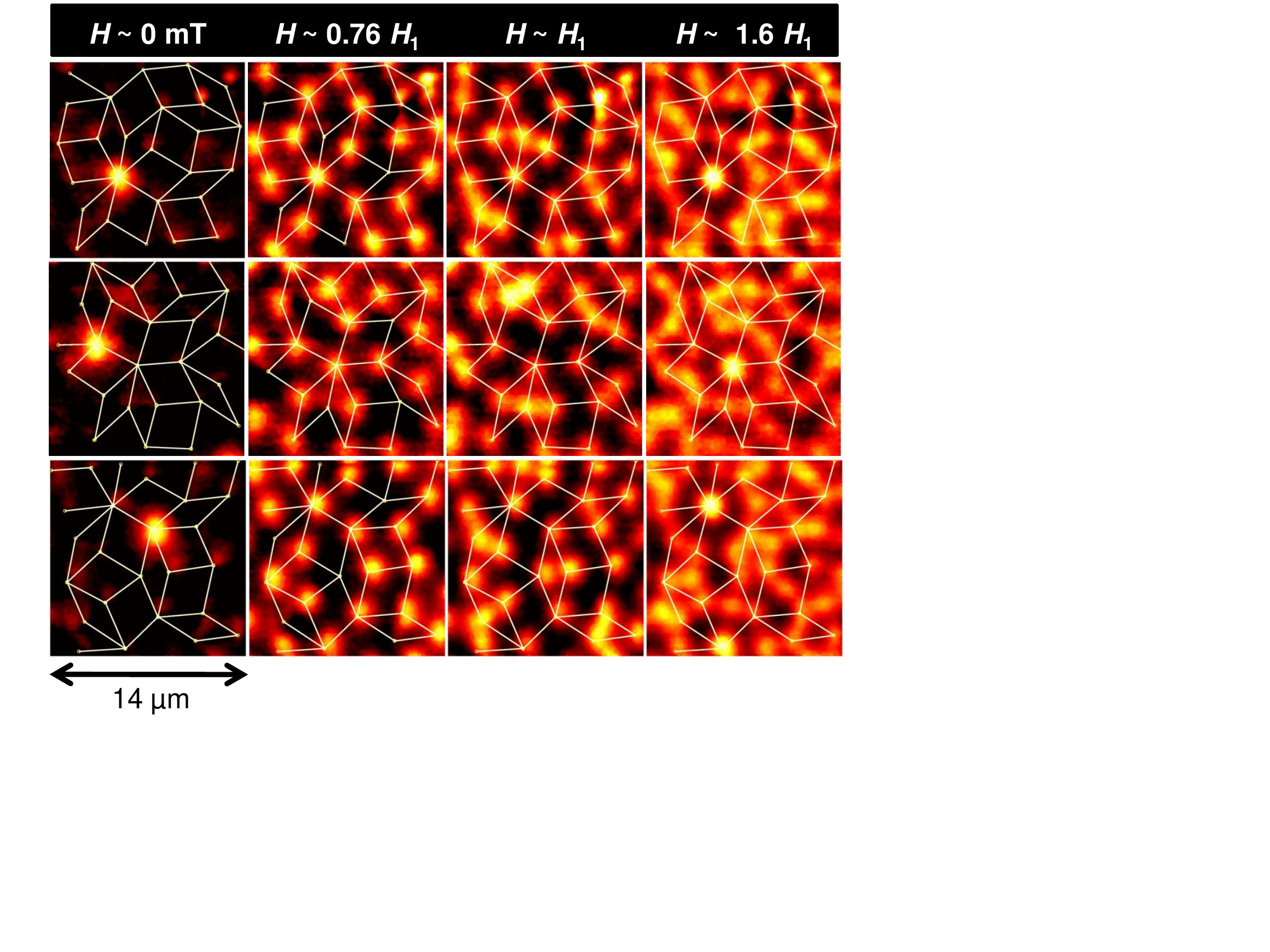}
\caption{(Color) Scanning Hall probe microscopy images obtained after field cooling the sample down to 4.2~K. Each row corresponds to a different location of the array. The first, second, third and fourth columns corresponds to $H \sim 0$~mT, $H \sim 0.220$~mT $\sim 0.76 H_1$, $H \sim 0.275$~mT $\sim H_1$, and $H \sim 0.455$~mT $\sim 1.6 H_1$, respectively.}
  \label{shpm}
\end{figure}

In order to clearly identify the microscopic vortex distribution associated with the different features observed in the $I_c(H)$ and $V(H)$ curves, we acquired scanning Hall probe microscopy images at different fields. A series of pictures obtained at 4.2~K after field cooling procedure with fields ranging from $-1$~mT to $+1$~mT in steps of 0.005~mT allowed us to determine the remanent field with high accuracy. Indeed, notice that steps of 0.010~mT correspond to one extra vortex per scanning area between two consecutive images. Figure~\ref{shpm} summarizes the obtained images for a selected set of fields~\cite{comment} (different columns) and at three different locations on the Penrose pattern (different rows). In this figure we indicate the Penrose tiles with white lines for clarity. The determination of the dots' position was achieved by performing a scan in presence of a large external field ($\sim 400$~mT) which produces a clear contrast of the Co microparticles due to their magnetic moment strongly aligned by the perpendicular field.

\begin{figure}[bt]
\centering
\includegraphics*[width=\linewidth, bb=170 112 519 384]{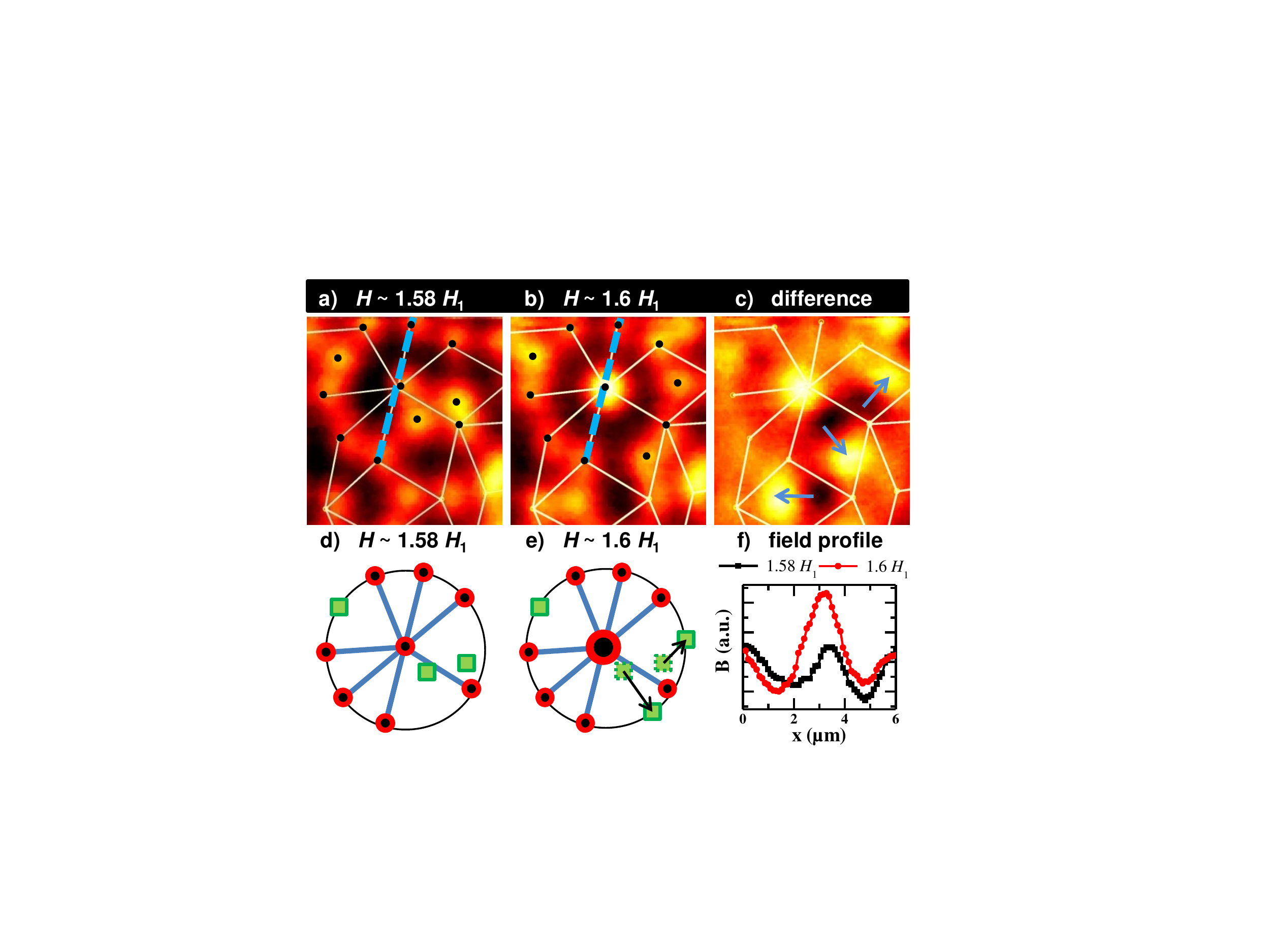}
\caption{(Color) Scanning Hall probe microscopy images obtained at two consecutive field values, $H \sim 0.445$~mT $\sim 1.58 H_1$ (a) and $H \sim 0.45$~mT $\sim 1.6 H_1$ (b) together with the differential image (c) taken at a specific location where the vortex corral appears. Black dots in (a) and (b) mark the vortices illustrated in (d) and (e). Black and white spots in panel (c) indicate the change of position of individual vortices. In panels (d) and (e) a schematic representation of the vortex distribution is given for the two consecutive field values. Red circles represent pinned vortices and green squares represent the interstitial vortices. Panel (f) shows the magnetic field profile obtained along the dotted line drawn in panel (a) and (b) for both magnetic field values. Integrated field for $1.6 H_1$ clearly corresponds to the giant $2\phi_0$-vortex.}
  \label{shpm2}
\end{figure}

The first column in Fig.~\ref{shpm} shows the local field profile obtained at $H \sim 0$. On this column, all panels show a single isolated vortex. The second column in Fig.~\ref{shpm} shows the vortex arrangement corresponding to the matching condition $H \sim 0.76 H_1$ (see also Fig.~\ref{transport}). It is clear from this image that unoccupied pinning sites are located at one of the vertices of the thin tiles. This vortex distribution was anticipated by Misko {\it et al.}~\cite{misko-PRB} based on the rapid increase of the vortex repulsion as the vortex separation decreases. The third column in Fig.~\ref{shpm} corresponds to the exact commensurability of the vortex lattice with the pinning landscape, i.e. every vortex sits on top of a Co dot. The last column illustrates the vortex pattern obtained at $H \sim 1.6 H_1$. Here, due to the repulsive interaction between vortices, interstitial vortices sit only inside the thick tiles. Differential images, obtained subtracting two consecutive field cooling, show that these interstitial vortices are very mobile and tend to avoid the geometrical center of the tile [Fig.~\ref{shpm2}(c)].

This finding indicates that an interstitial bistable configuration takes place inside each thick tile. The formation of ring-like structures or vortex corrals clearly seen in the last column of Fig.~\ref{shpm}, demonstrates that a long range interaction eventually lifts the two-fold degeneracy of the interstitial positions. More importantly, a detailed analysis of the field intensity at the core of the ring-shaped vortex structures indicates that the central vortex carries {\it two flux quanta} $\phi_0$ [see Fig.~\ref{shpm2}(f)]. These multiquanta vortices are not a mere consequence of stochastic distribution of pinning strength since they appear at well defined locations on the pinning landscape corresponding to multinodal agglomeration of Penrose tiles.

To obtain a better insight in the mechanism responsible for the creation of these multiquanta vortices we show the field profile corresponding to two consecutive images and their difference [Fig.~\ref{shpm2} panel (a), (b) and (c)] exactly at the creation of a vortex corral ($H \sim 1.6 H_1$). A reconfiguration of the interstitial vortices creates a highly symmetrical vortex structure around specific multinodal points of the Penrose lattice. This process is schematically presented in Fig.~\ref{shpm2} [panel (d) and (e)], which shows the distribution of pinned (red circles) and interstitial (green squares) vortices for both consecutive magnetic fields. At $H \sim 1.6 H_1$ the mobile interstitial vortices rearrange themselves in a structure with high rotational order. The exact locations, where the vortex corral appears, are surrounded by ten vortices, seven pinned at the Penrose lattice and three interstitial vortices. As a result a ten-fold rotational symmetry is obtained [Fig.~\ref{shpm2}(b) and (e)]. At the center of this highly symmetric vortex structure a multiquanta vortex is favored.

This unexpected result resembles the symmetry-induced giant vortex states in mesoscopic structures close to the superconducting/normal metal phase boundary~\cite{moshchalkov,bruyndoncx,kanda}. In that case, the boundary conditions impose the sample symmetry on the order parameter which manifest itself in the formation of unconventional vortex patters as giant vortices or vortex-antivortex pairs. Here, the combination of the applied magnetic pressure, induced by the surrounding vortices, and the high local symmetry leads to the generation of a multiquanta vortex at the center of the vortex corral. This is in agreement with previous experimental~\cite{moshchalkov, bezryadin} and theoretical~\cite{doria} reports showing that the number of vortices trapped by a single pinning site increases with increasing external field. In our particular case, it is the highly inhomogeneous {\it local field} which drives the system to the multiquanta vortex state.

It is important to emphasize that the theoretical investigations of the vortex matter in a five-fold Penrose array performed so far~\cite{misko-PRB,misko-PRL}, assumed a short range interaction between vortices. Since the present work has been done on a thin Pb film with thickness smaller than the temperature dependent penetration depth, a more suitable comparison should be done with long range repulsive Pearl vortices in the presence of a quasiperiodic array of pinning sites~\cite{pearl}. The combination of this long range vortex-vortex interaction with the large penetration depth expected by freezing the flux lattice at high temperatures are most likely necessary ingredients for the formation of a giant vortex state at $H \sim 1.6 H_1$.


In conclusion, we have investigated the vortex distribution in a superconducting film with a Penrose lattice of Co dots by scanning Hall probe microscopy. The obtained images confirmed the theoretically predicted vortex patterns at low densities (i.e. $H \leq H_1$). A much richer vortex structure is found above $H_1$ where mobile interstitial vortices create highly symmetrical vortex corrals around particular pinning sites of the Penrose lattice. The formation of vortex rings in combination with long range vortex-vortex interaction leads to the stabilization of multiquanta vortices at the core of these rings.


This work was supported by Methusalem Funding of the Flemish government, FWO-Vlaanderen, the Belgian Inter-University Attraction Poles IAP, and the ESF-NES Programmes. A.V.S. and J.V.d.V are grateful for the support from the FWO-Vlaanderen.

\end{document}